\documentclass[11pt]{article}
\usepackage{fullpage}
\usepackage{comment}
\usepackage[hyphens]{url}
\usepackage{graphicx}
\usepackage{amsmath}
\usepackage{amsfonts}
\usepackage{float}
\title{\LARGE{Detecting Pump\&Dump Stock Market Manipulation from Online Forums}}
\author{D. Nam $\;\;\;$ D.B. Skillicorn\\
School of Computing \\
Queen's University \\
Kingston.  Canada \\
skill@queensu.ca}
\date{}
\begin{document}
\maketitle
\bibliographystyle{plain}

\begin{abstract}
{\sf
The intersection of social media, low-cost trading platforms, and
naive investors has created an ideal situation for information-based
market manipulations, especially pump\&dumps. Manipulators
accumulate small-cap stocks, disseminate false information on social
media to inflate their price, and sell at the peak.
We collect a dataset of stocks whose price and volume profiles
have the characteristic shape of a pump\&dump, 
and social media posts for those same stocks that match the timing of the 
initial price rises.
From these we build predictive models for pump\&dump
events based on the language used in the social media posts.

There are multiple difficulties: not every post will cause the
intended market reaction, some pump\&dump events may be triggered by
posts in other forums, and there may be accidental confluences of
post timing and market movements.
Nevertheless, our best model achieves a prediction accuracy of 85\%
and an F1-score of 62\%.
Such a tool can provide early warning to investors and regulators
that a pump\&dump may be underway.
}
\end{abstract}

\section{Introduction}

New financial products and technologies have allowed 
naive investors to easily enter financial markets. This has increased the 
risk of manipulation, and detecting and investigating fraudulent activities 
has become much more difficult. Many go undetected \cite{forde_2013}. 

Social media has created new methods for manipulating markets. 
A scheme known as \emph{Pump and Dump} (P\&D) is one popular mechanism. 
Fraudsters buy quantities of a stock, disseminate false information about 
it to artificially raise its price, and then sell their purchased shares 
at the higher price. 
Social media provides a channel for rapid dissemination and a pool of
investors with little knowledge or experience who may not detect
that the information is false.

Conventional approaches to detecting manipulation 
look for known patterns, and for anomalous activity such as
exceeded thresholds for prices and trading volumes. 
Suspicious activities can be detected using sets of rules 
and triggers that cause notifications of potential manipulation.
However, those methods struggle in the presence of behaviours that deviate 
from historical patterns \cite{golmohammadi_detecting_2014}.
Previous work has also focused on detecting manipulations so that regulators
can penalise those who carry them out. 
This does little to help investors, either
to prevent their being deceived or recovering their investments.

Data-analytic techniques have the potential to detect false information
as it being disseminated \cite{delort_2011, owda_financial_2017}.
Natural language analytics can detect the posts in social media
that are intended to pump particular stocks, providing a real-time
warning to potential investors.
We investigate how well P\&D schemes can be detected
in posts on social media, by matching the language patterns in
the posts to the pattern of stock price corresponding to a
P\&D manipulation.

A penny stock is a stock that is traded by a small public company for 
less than \$5 per share \cite{murphy_2020}. Many of these companies are 
known for their volatility due to their limited coverage by analysts and 
interest from institutional buyers. 
Because of their low price, retail investors can buy large quantities 
of these stocks without having to invest much money. 
This, however, makes their prices volatile and so creates the 
potential for large returns on investments; but also leaves them
vulnerable to manipulation by malicious actors. One study found that 
50\% of manipulated stocks are those with a small market 
capitalization \cite{aggarwal_stock_2006}.   

It might be supposed that the connection between a social media
post and a P\&D event is too tenuous to be detected --
after all, not every post will have the desired effect, and a
P\&D might be triggered by some less visible social media
activity. 
We show that, at least for penny stocks, the connection
is reasonably detectable, and we achieve prediction accuracies
(that a post is intended to cause a P\&D event) of 85\%, with
an F1 score of 67\% ($\pm$ 12 percentage points) from posts
alone, and 62\% ($\pm$ 3 percentage points) from posts and comments.

\section{Tools}

Stance detection is a technique to determine the attitude or viewpoint 
of a text towards a target. 
It aims to detect whether the author of the text is in support of or 
against a given entity \cite{li-caragea-2019-multi}. 
Some applications of stance detection have been in political debates, 
fake news, 
and social media \cite{ghanem-etal-2018-stance, rajaash,thomas-etal-2006-get}.  

Empath is a tool that was developed by Fast et al. \cite{fast_empath_2016} 
for researchers to generate and validate new lexical categories on demand. 
It uses deep learning to establish connections between words and phrases
used in modern fiction. Given a small set of seed words that represents 
a category, Empath can provide new related terms using its neural embeddings. 
It also employs the use of crowd-sourcing to validate the terms that it
considers are related. 
Along with the ability to create new categories, Empath comes with 200 
built-in, pre-validated categories for common topics (e.g., neglect, 
government, social media).  

SHAP (\textbf{SH}apley \textbf{A}dditive ex\textbf{P}lanation) is a 
tool that was developed by Lundberg and Lee \cite{NIPS2017_7062}
to determine the impact of each attribute
on the output of a predictive model.
It is based on Shapley values, a concept from game theory that determines 
a fair way to distribute the payoff for players that have worked
in coalition towards an outcome \cite{wild_2018}. 

Extreme Gradient Boosting is a decision-tree based ensemble algorithm that 
has become known for its speed and performance \cite{brownlee_2020}. 
Decision trees are built sequentially so that each one reduces the errors 
of the previous one \cite{vidhya_2020}.
Random Forests is a decision-tree based ensemble algorithm with
each tree built from a subset of the rows and columns of
the dataset \cite{wood_2020}. This allows for variation among the trees 
and results in lower correlation among their predictions \cite{yiu_2019}.
Support Vector Machines are a supervised learning algorithm that finds 
a hyperplane that best separates the data points from two 
classes \cite{gandhi_2018}.

Artificial Neural Networks are computational networks that are inspired 
by the biological nervous system \cite{dacombe_2017}.
ANNs excel at prediction for data where the amount of information
in each attribute is small and there are non-linear interactions
among them.
Deep learning models are a class of extensions to ANNs that have
solved long standing prediction problems in image recognition
and natural language \cite{lecun_deep_2015}. 
Convolutional Neural Networks (CNNs) are a class of deep learning
networks that were designed initially to work with 
images but work surprisingly well with sequence data such
as texts as well.
Long Short-Term Memory (LSTM) deep learning networks are
a type of recurrent neural network designed to handle the
long-term dependencies present in sequence prediction problems 
\cite{brownlee_2020_lstm}. Understanding text
often requires looking ahead (think of verbs in German)
and so processing text in both directions, using a biLSTM,
provides better results for language \cite{brownlee_2020_bilstm}. 

\section{Experiments}\label{ch:Experiments}

Within a typical online forum, there are two different categories of texts. 
The first is a \emph{post}, which initiates a discussion. 
The second is a set of \emph{comments} responding to the post. 
For example, an individual may post saying that, in their opinion, 
a stock's price is about to rise, with others respond by
sharing their opinions in the same thread. 
Responders may agree with the original post, or disagree.

\begin{figure}[t]
  \centering
    \includegraphics[width=0.80\textwidth]{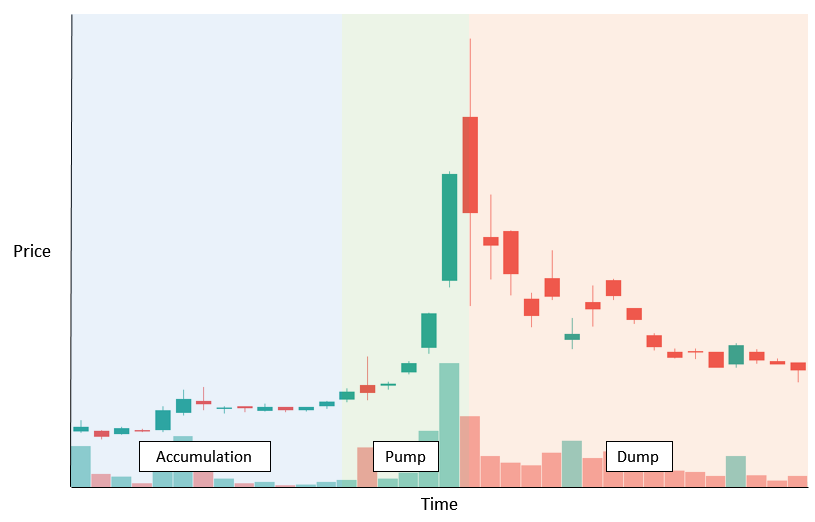}
    \caption{Stages of Pump and Dump}
    \label{pnd}
\end{figure}

P\&D is an information-based manipulation, artificially raising the 
price of a stock through the dissemination of false information. 
As shown in Figure \ref{pnd}, this manipulation strategy involves 
three different stages \cite{kamps18}. 
The operators of the scheme first purchase the stock that they are planning 
to manipulate (Accumulation). Once they have acquired enough shares, 
they will disseminate false information to make it appear more desirable, 
driving up the price (Pump). Once the price has risen to the desired 
level of profit, the operators sell off their shares before anyone 
uncovers that the information has no basis or the hype dies down (Dump).

To identify P\&Ds within the market, patterns associated with
the scheme must be established. 
While the method of conducting a P\&D may vary, 
two indicators that can identify them are sharp
changes in price 
and volume \cite{kamps18}. A P\&D will cause a significant price
increase within a short amount of time, larger than the fluctuations 
that the stock typically experiences; followed by 
a decrease once the dump phase has begun. 
The volume also increases as the stock gains interest 
among investors during and after the dissemination phase. 
However, the volume will typically not immediately experience as sharp
a decline as the price when the operators begin to dump their shares
because of the reluctance of investors to believe that the price
is illusory.

If the profile of a P\&D manipulation can be detected in
the market, then the post that putatively caused it can be straightforwardly
labelled and its language patterns investigated.
(Of course, it is possible that some of the apparent connections
are spurious, but it is relatively unlikely that a post touting
a particular stock will be disseminated exactly when the stock's
price and volume begin a sharp rise).

Labelling comments is more complex, since the comments may agree
with the original post, or disagree. Only the language of those
that agree can contribute to predicting a P\&D event.

\subsection{Data Sources}

Two different data sources were utilized. The first is the popular 
online website Reddit, where users discuss the stock market. 
The second is Yahoo Finance, a financial market 
website that provides historical data about companies.

Reddit contains forums referred to as subreddits, each dedicated to the 
discussion of a specific topic. 
Popular forums for the discussions of stocks are r/pennystocks,
r/wallstreetbets, r/stocks, r/RobinHoodPennyStocks, r/TheWallStreet. 
We use r/pennystocks and r/RobinHoodPennyStocks,
 
Yahoo Finance is a website provided by Yahoo for investors to 
access financial news, market data, and basic financial tools.
Given a stock symbol or company name, it provides the relevant market
data.

\begin{figure}[h]
  \centering
    \includegraphics[width=1.0\textwidth]{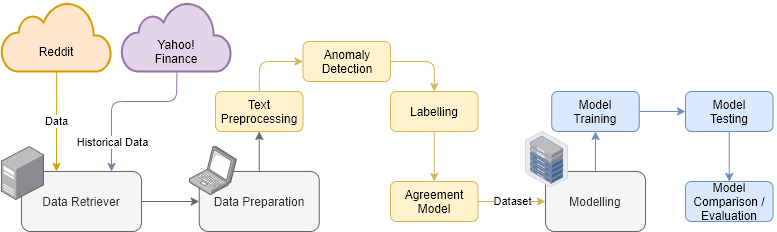}
    \caption{Experiment workflow}
    \label{experimentsetup}
\end{figure}

Classification techniques such as Extreme Gradient Boosting (XGBoost), 
Random Forests, Support Vector Machine (SVM), and Artificial Neural Networks 
(ANNs) were used to learn predictive models, and then to identify which
attributes (i.e. words) are most predictive.
Figure \ref{experimentsetup} shows the experimental workflow.

Data from Reddit and Yahoo Finance were collected daily
for the period October 1, 2019, to June 28, 2020. 
A breakdown of the data is shown 
in Table~\ref{databreakdown}. The majority of the data is retrieved 
from r/pennystocks, with about a third from r/RobinHoodPennyStocks. 
The number of comments is much larger than the number of posts,
with posts making up only about 5\% of the texts. 

\begin{figure}[t]
  \centering
    \includegraphics[width=0.9\textwidth]{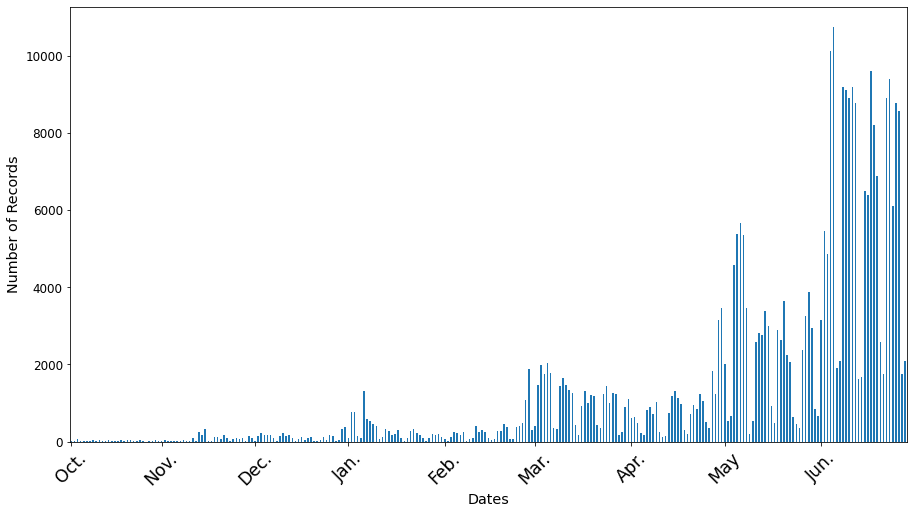}
    \caption{Data Collection Volumes}
    \label{datatrend}
\end{figure}

\begin{table}[t]
\centering
\resizebox{1.0\textwidth}{!}{
\begin{tabular}{|p{5cm}|p{5cm}|p{5cm}|p{1.5cm}|}
\hline
\textbf{Subreddit}              & \textbf{Number of Posts} & \textbf{Number of Comments} & \textbf{Total} \\ \hline
r/pennystocks          & 12,049             & 234,149  &   246,198          \\ \hline
r/RobinHoodPennyStocks & 6,506             & 78,429    &  84,935         \\ \hline
\textbf{Total}          & 18,555       &  312,578  & 331,133    \\ \hline
\end{tabular}}
\caption{Breakdown of records collected from subreddits}
\label{databreakdown}
\end{table}
 As shown in Figure \ref{datatrend}, there was a sharp increase
in the number of submissions over the period of data collection:
\begin{itemize}
    \item r/pennystocks - 139,000 Members $\Rightarrow$ 257,000 Members
    \item r/RobinHoodPennyStocks - 52,000 Members $\Rightarrow$ 133,0000 Members
\end{itemize}
This seems to reflect an increase in amateur stock market investing
because of the covid-19 pandemic, and a corresponding increase
in manipulation. i.e, as manipulators look to take advantage of 
new, naive investors during the pandemic. 
Alerts and press releases by the SEC and the Canadian 
Securities Administrators warned new investors to be vigilant 
about the increasing number of P\&D schemes that have occurred around
that time \cite{csa_2020, sec_2020, sec2_2020}.

The median number of words per post or comment was 22, and the total
number of distinct words was 4,862.

\begin{figure}[t]
  \centering
    \includegraphics[width=0.8\textwidth]{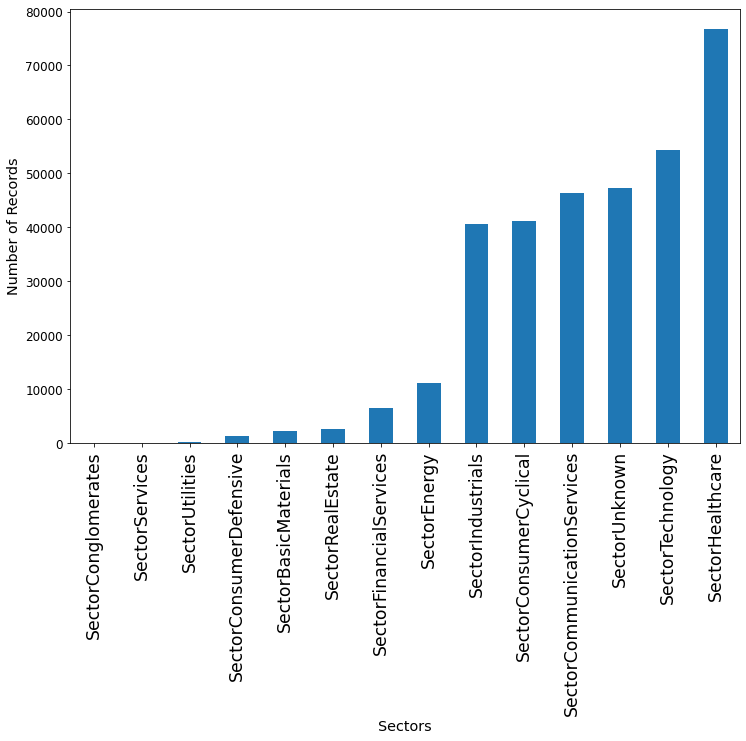}
    \caption{Histogram of market sectors discussed within subreddits}
    \label{stocksectors}
\end{figure}

\begin{figure}[t]
  \centering
    \includegraphics[width=0.9\textwidth]{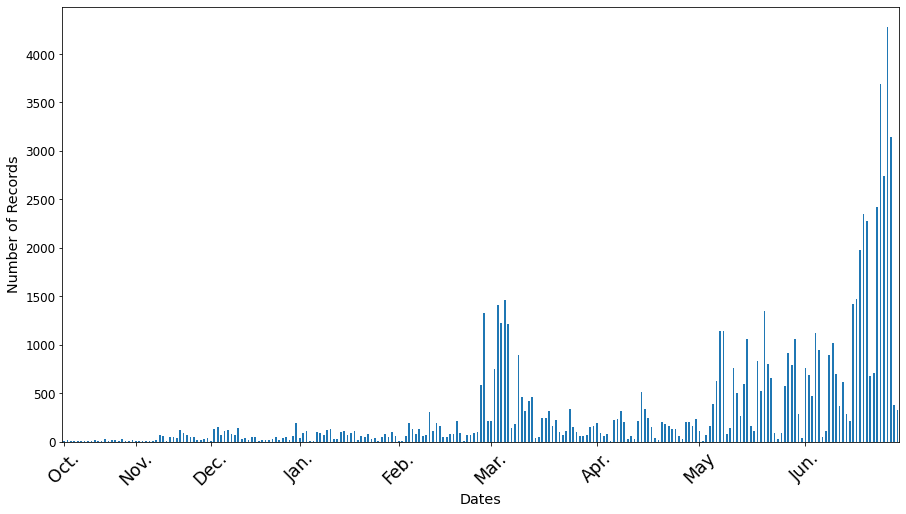}
    \caption[Healthcare posts and comments trend]{Trend of posts and comments that discussed healthcare stocks}
    \label{healthcaretrend}
\end{figure}

\begin{figure}[t]
  \centering
    \includegraphics[width=0.9\textwidth]{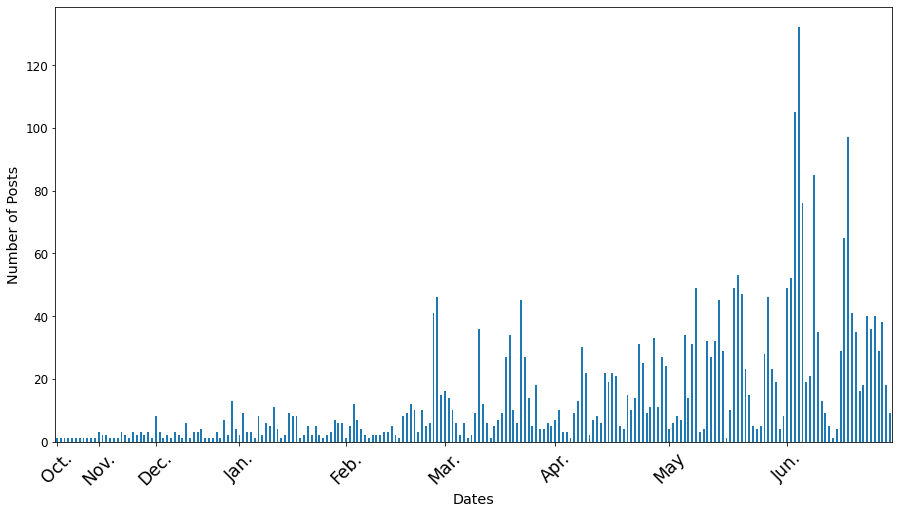}
    \caption[Pump and dump posts trend]{Trend of posts that have been labelled as P\&D}
    \label{manipulation_trend}
\end{figure}

Replacing stock symbols by the market sector to which each business
belongs allows us to see which sectors are discussed the most,
and which are the targets of P\&D.
Figure \ref{stocksectors} shows that healthcare stocks are the most 
mentioned, followed by technology stocks. 
The pandemic clearly had an effect on both attention to markets
and manipulations.
Temporal trends in the healthcare sector,
Figure~\ref{healthcaretrend} ,
show an increase in online activity at the beginning of
the pandemic, and then a further increase in the middle of 2020.
Figure~\ref{manipulation_trend} shows that P\&D manipulations
also increased in 2020.

Table \ref{postcommentdatafeature} shows the information collected 
for each post and comment.

\begin{table}[t]
\centering
\resizebox{0.90\textwidth}{!}{
\begin{tabular}{p{5cm}p{8cm}}
\hline
\textbf{Feature} & \textbf{Description}  \\ \hline
{Post Title} & {Title of the post.}  \\ \hline
{Post ID} & {Unique identification code for post.}  \\ \hline
{Post Author} & {Author of the post.}  \\ \hline
{Post Created} & {Unix Timestamp of when post was submitted.}  \\ \hline
{Post Body} & {Text of the post.}  \\ \hline
{Comment ID} & {Unique identification code for comment.}  \\ \hline
{Comment Author} & {Author of the comment.}  \\ \hline
{Comment Created} & {Unix Timestamp of when comment was submitted.}  \\ \hline
{Comment Body} & {Text of the comment.}  \\ \hline
\end{tabular}}
\caption{Features of collected Reddit data}
\label{postcommentdatafeature}
\end{table}

Data from Yahoo Finance was scraped using the yfinance tool \cite{yfinance}. 
Stock symbols were extracted from Reddit posts. This step is non-trivial
and required regular expression extraction, and look ups against the publicly
traded exchanges.
Posts which mentioned more than one stock were discarded, partly because
of the complexity of deciding which stock may be being touted, and partly
because P\&D posts typically focus on one particular stock they are
pumping.
If a stock symbol was found, yfinance was used to collect the financial 
information described in Table \ref{yahoofinancedatafeature}. 

\begin{table}[t]
\centering
\resizebox{0.90\textwidth}{!}{
\begin{tabular}{p{5cm}p{8cm}}
\hline
\textbf{Feature} & \textbf{Description}  \\ \hline
{Open} & {Opening price of the stock for the given period.}  \\ \hline
{High} & {Highest price for the stock within the given period.}  \\ \hline
{Low} & {Lowest price for the stock within the given period.}  \\ \hline
{Close} & {Closing price of the stock for the given period.}  \\ \hline
{Volume} & {Total number of shares traded within the given period.}  \\ \hline
{Market Sector} & {Associated industry that the company is in.}  \\ \hline
{Market Capitalization} & {Total market value of the company's outstanding shares.}  \\ \hline
\end{tabular}}
\caption{Features of Yahoo! Finance data}
\label{yahoofinancedatafeature}
\end{table}

\begin{figure}[t]
  \centering
    \includegraphics[width=0.8\textwidth]{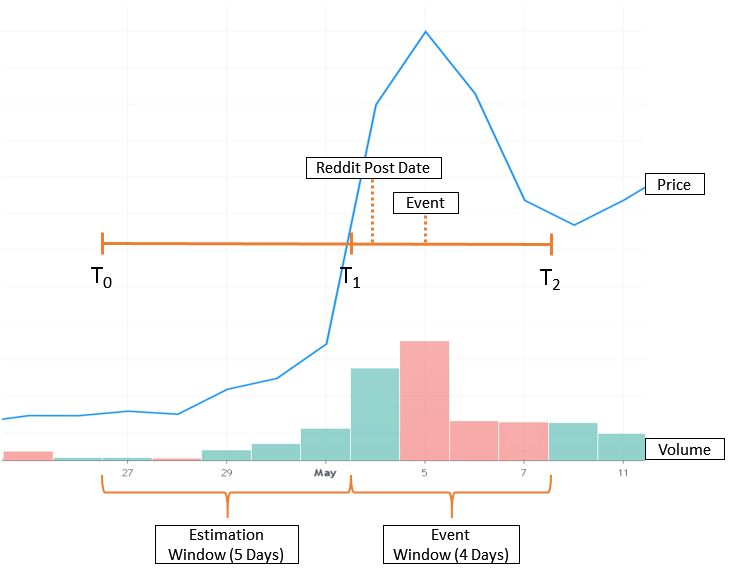}
    \caption{Time window used to collect market data.}
    \label{stockwindow}
\end{figure}

As shown in Figure \ref{stockwindow}, the daily Open, High, Low, Close, 
and Volume (OHLCV) data was collected over nine business days surrounding
an event. 
Data was collected over five days before each post event to
establish a baseline for price and volume. Penny stocks
almost always shows minor variation in price and volume so this baseline is
typically quite flat.
The remaining four days contain the pump event (sharp increase)
followed by a decrease in price and a slower decrease in volume.
Sabherwal et al. \cite{sabherwal_internet_2011} studied the effects of 
online message boards on market manipulation and found that dumps
typically occur within four days and this is plausible because the
manipulators want to sell as soon as the price reaches a peak.

Texts from subreddits were preprocessed using the following steps:
remove URLs,
expand contractions,
remove HTML Tags,
remove punctuation,
remove extra whitespaces,
remove numbers,
lemmatization, and
remove stopwords.

Stock symbols within the text were replaced by dummy stock names
representing the market sector associated with each business.
This is required because the name of the particular stock being
pumped and dumped in one case has nothing to do with the name of
the stock being used in another case -- but there might be correspondences
within sectors.
Here is an example:
\begin{itemize}
    \item ``\textbf{AYTU} perfect time to buy" $\Rightarrow$ ``\textbf{SectorHealthcare} perfect time to buy"
\end{itemize}

\subsection{Data Labelling}

To label each post, stock data surrounding the day in which the post was 
submitted to Reddit were analyzed. 
If the market data exhibited that pattern associated with P\&D (a notable
rise from the time of the post, followed by a sharp drop) then the post
was labelled accordingly.
A rise was detected by calculating the average price and volume in the
five-day window before the post.
The daily average price (\textbf{DAP}) of the values was first 
calculated for each of the five days.   
\begin{equation}
    DAP(X_t) = \frac{1}{4}(X_{t_{open}} + X_{t_{high}} + X_{t_{low}} + X_{t_{close}})
    \label{dailyaverage}
\end{equation}
and then the baseline average price (\textbf{BAP}) was calculated by
\begin{equation}
    BAP(X_{est}) = \frac{1}{5}\cdot\sum_{t=T_{0}}^{T_{1}} DAP(X_t)
    \label{baselineaverageprice}
\end{equation}
The baseline average volume (\textbf{BAV}) was calculated by taking 
the average of the volume values over the estimation window.
\begin{equation}
    BAV(X_{est}) = \frac{1}{5}\cdot\sum_{t=T_{0}}^{T_{1}} X_{t_{volume}}
    \label{baselineaveragevolume}
\end{equation}

\begin{figure}[t]
  \centering
    \includegraphics[width=0.9\textwidth]{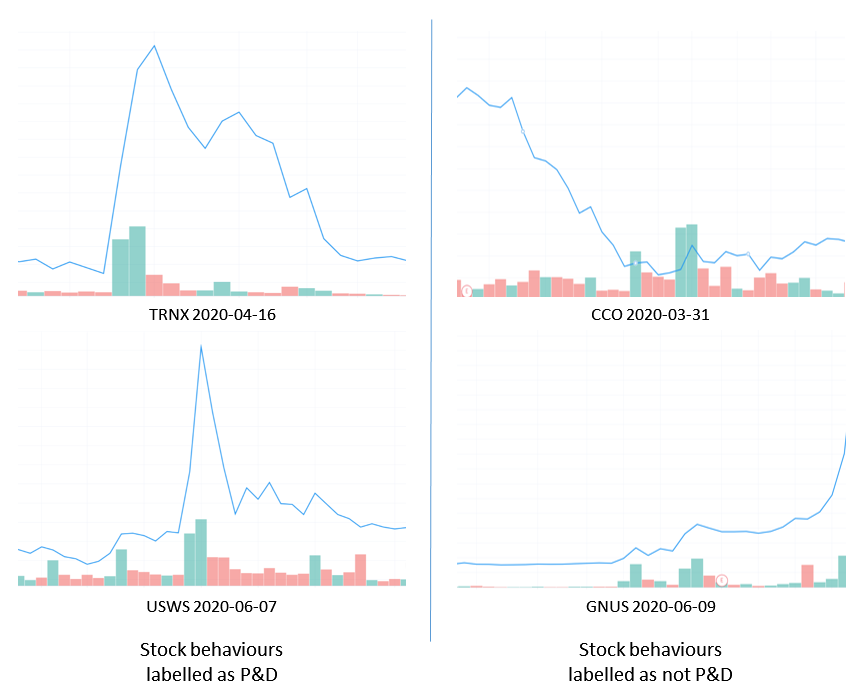}
    \caption[Labelling of stock behaviours]{Comparison of stock behaviours that have been labelled using anomaly detection}
    \label{anomalycomparison}
\end{figure}

A threshold was set at two standard deviations above the average price 
within the five-day estimation window. Price increases above this threshold
were considered to be pump events.
A similar threshold was used to define a volume anomaly.
Events were considered to be the result of P\&D if they exceeded the
threshold for both price and volume.
Figure \ref{anomalycomparison} shows a comparison of the stock behaviours
labelled using this approach. 

\begin{figure}[t]
  \centering
    \includegraphics[width=0.9\textwidth]{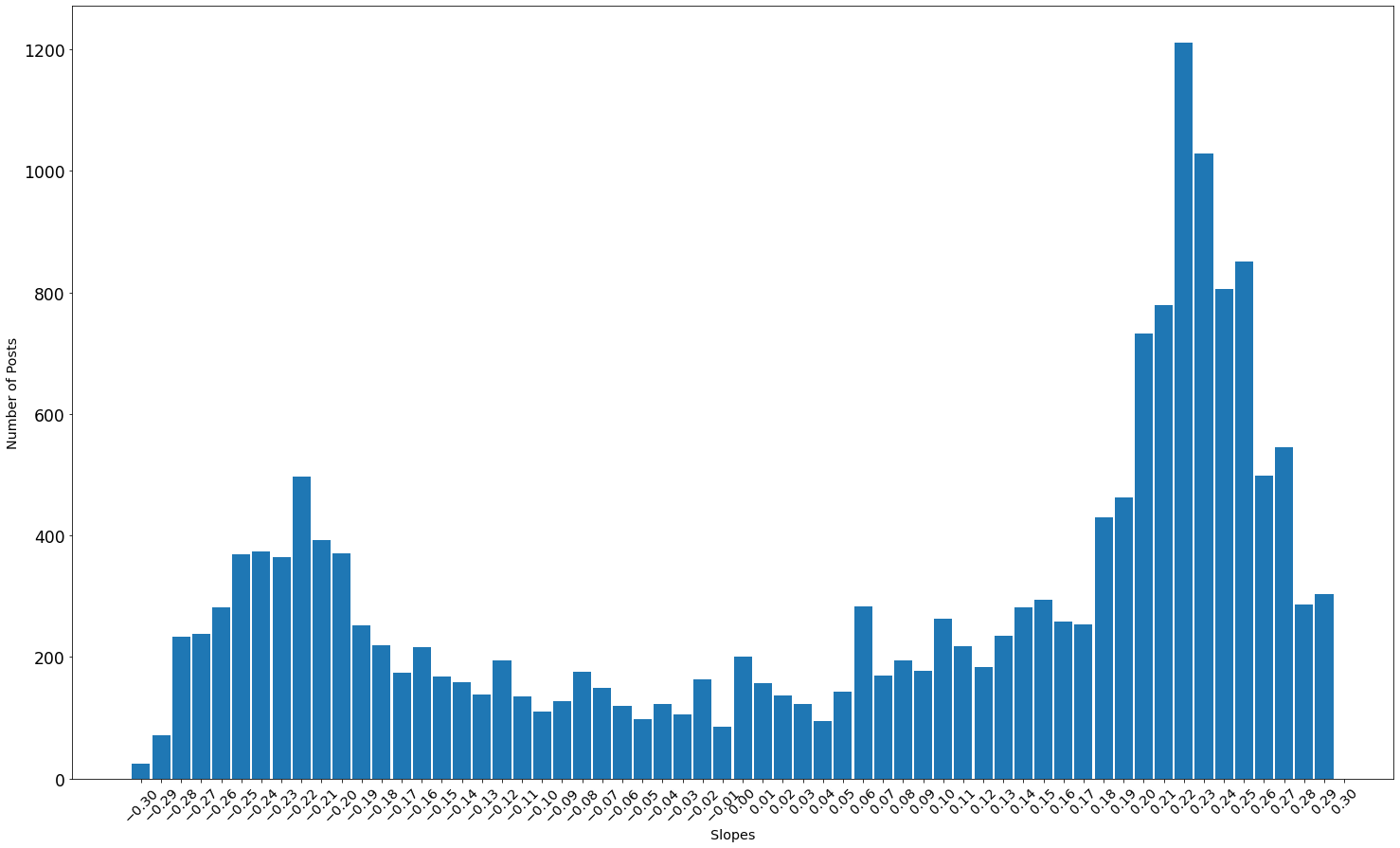}
    \caption{Distribution of stock price trend slopes}
    \label{pricetrend}
\end{figure}

A sudden price rise or volume increase might coincide with a post,
but is not necessarily caused by it.
The rising region of each stock trend of a potential P\&D event
was min-max normalised, and its slope calculated.
Steep price increases are more likely to arise from genuine
information and less likely to have resulted from
a single manipulation post, so the median slope across the entire dataset
was calculated, and only slopes below the median were considered
as potential P\&D events.
Figure~\ref{pricetrend} shows the distribution of stock price trend slopes 
from the entire the dataset. The median value is 0.18.

\subsection{Agreement Model}

The comments associated with the P\&D post cannot all be labelled
as examples of P\&D language, since not all of them will be
supportive of the post they are responding to.
Manipulators, of course, will post comments in support of the
post, either from the same identity or from others.

We developed an agreement model, using ideas from stance detection.
This was done using Empath to generate a lexicon of agreement,
seeding it with the words: \textbf{bought}, \textbf{agree}, \textbf{positive}, \textbf{increasing}, \textbf{good}, and \textbf{now}. Empath returned the 
words listed in Table~\ref{empathwords}.
\begin{table}[t]
\centering
\resizebox{0.85\textwidth}{!}{
\begin{tabular}{|*{5}{p{2cm}|}}
\hline
only & done & better & true & knew\\
besides & like & maybe & wanted & liked\\
also & important & buying & understand & good\\
understood & needed & work & because & successful\\
knowing & grateful & plus & much & reasonable\\
should & give & happy & course & glad\\
well & considering & anyway & agree & meaning\\
great & probably & sure & thought & guaranteed\\
more & honestly & positive & thankful & actually\\
agreed & special & doubt & guess & though\\
bet & buy & surpass & worth & \\
suppose & although & especially & definitely & \\
certain & figured & given & means & \\
\hline
\end{tabular}}
\caption{List of generated agreement words from Empath}
\label{empathwords}
\end{table}
Posts touting stocks also use a specialised vocabulary, shown in these
examples. 
\begin{itemize}
    \item ``probably go to \textbf{shoot} up tomorrow"
    \item ``this bad boy just \textbf{rocket}"
    \item ``i will see you on the \textbf{moon}"
\end{itemize}
An extended lexicon was determined manually by inspecting posts associated 
with manipulation.
Table~\ref{customwords} contains the list of words that were chosen using
this approach.
\begin{table}[t]
\centering
\resizebox{0.85\textwidth}{!}{
\begin{tabular}{|*{5}{p{2cm}|}}
\hline
moon & fast & massive & rich & surprise\\
rocket & profit & top & easy & move\\
pump & rally & peak & early & load\\
soar & climb & worth & shoot & quick\\
jump & rise & sale & money & burst\\
pop & high & gain & breakout & drive\\
hype & spike & run & cash & nice\\
fly & go & up & hit & bank \\
awesome & confident & surpass & more & zoom\\
big & great & potential & advantage & \\
\hline
\end{tabular}}
\caption{List of custom words used in the Agreement Model}
\label{customwords}
\end{table}

Comments were labelled as associated with pumping if they contained
two or more of the agreement words, or if they were (visibly) authored by
the original poster.
The following are some examples of comments that were labelled as not 
P\&D related based on the agreement model:
\begin{itemize}
    \item ``it be the american dream to fall for snake oil salesman and then lose everything it be a story as old as humanity"
    \item ``clearly a pump and dump scheme"
    \item ``do not touch it if the chart look like a hockey stick"
\end{itemize}
This labelling of comments is limited by the completeness
of the agreement lexicon, and also does not account for negations.

P\&D posts and comments are relatively rare and so the dataset 
is naturally imbalanced. Techniques such as SMOTE \cite{smote} 
and ADASYN \cite{adasyn} were tried but proved ineffective.
Instead, where predictors allowed it, class weight parameters
were set to penalise mistakes in the minority class.

\subsection{Modelling}

The following predictors were used:
\begin{itemize}
    \item Extreme Gradient Boosting (XGBoost)
    \item Random Forest (RF) 
    \item Support Vector Machine (SVM)
    \item Artificial Neural Networks
    \begin{itemize}
        \item Multilayer Perceptron (MLP)
        \item Convolutional Neural Network (CNN)
        \item Bidirectional Long Short Term Memory (BiLSTM)
    \end{itemize}
\end{itemize}

In each case the standard performance measures (accuracy, precision, recall, 
F1-Score, confusion matrix) were calculated, as well as the
Shapley values which rank words by their importance to the
predictions.

\section{Results}\label{ch:Results}

Table \ref{classdistribution} shows the class distribution for the dataset. 
Less than 9\% of the records are labelled as being P\&D. This is typical of
datasets where fraud is present; indeed it is striking that the rate
of fraud is this high.

\begin{table}[t]
\centering
\resizebox{1.0\textwidth}{!}{
\begin{tabular}{|p{3cm}|p{4cm}|p{5cm}|p{1.5cm}|}
\hline
\textbf{Record Type} & \textbf{P\&D} & \textbf{Not P\&D} & \textbf{Total} \\ \hline
Posts    &     3,006            &       15,549         &   18,555   \\ \hline
Comment &      26,727             &  285,851         & 312,578    \\ \hline
\textbf{Total}    &        29,733         &     312,142        &    331,133    \\ \hline
\end{tabular}}
\caption{Dataset class distribution}
\label{classdistribution}
\end{table}

The results of each of the predictive model are reported in
Table~\ref{modelperformance} using 5-fold cross validation and
upweighting the fraud class when the model permits it.

\begin{table}[t]
\centering
\scalebox{0.7}{
\begin{tabular}{|c|c|c|c|c|c|c|c|c|}
\hline
\textbf{Model} &
  \textbf{TP} &
  \textbf{FP} &
  \textbf{TN} &
  \textbf{FN} &
  \textbf{Accuracy} &
  \textbf{Precision} &
  \textbf{Recall} &
  \textbf{F1-Score} \\ \hline\hline
\textbf{XGBoost Posts} &
   1728 &
   6615 &
   8934 &
   1278 &
  57.46 ($\pm$3.73) &
  20.71 ($\pm$0.48) &
  57.49 ($\pm$0.68) &
  30.45 ($\pm$2.25) \\
 \hline
\textbf{XGBoost Posts and Comments} &
   2007 &
   7646 &
   7903 &
   999 &
  53.41 ($\pm$1.42) &
  20.79 ($\pm$0.85) &
  66.77 ($\pm$1.58) &
  31.71 ($\pm$0.96) \\
 \hline\hline
\textbf{RF Posts} &
   271 &
   646 &
   14903 &
   2735 &
  81.78 ($\pm$0.51) &
  29.55 ($\pm$1.40) &
  9.01 ($\pm$0.52) &
  13.81 ($\pm$0.78) \\
 \hline
\textbf{RF Posts and Comments} &
   414 &
   211 &
   15338 &
   2592 &
  84.89 ($\pm$0.69) &
  66.24 ($\pm$1.69) &
  13.77 ($\pm$0.47) &
  22.80 ($\pm$0.75) \\
 \hline\hline
\textbf{SVM Posts} &
   1752 &
   5263 &
   10286 &
   1254 &
  64.88 ($\pm$1.14) &
  24.98 ($\pm$0.76) &
  58.28 ($\pm$1.05) &
  34.97 ($\pm$1.16) \\
  \hline
\textbf{SVM Posts and Comments} &
   2125 &
   4559 &
   10990 &
   881 &
   70.6 ($\pm$0.49) &
  31.79 ($\pm$0.43) &
  70.69 ($\pm$0.56) &
  43.86 ($\pm$0.57) \\ 
 \hline\hline
\textbf{MLP Posts} &
   2382 &
   1718 &
   13831 &
   624 &
  87.38 ($\pm$6.66) &
  58.10 ($\pm$11.65) &
  79.24 ($\pm$12.76) &
  67.04 ($\pm$12.12) \\
 \hline
\textbf{MLP Posts and Comments} &
   2103 &
   2602 &
   12947 &
   903 &
  81.11 ($\pm$3.71) &
  44.70 ($\pm$4.28) &
  69.96 ($\pm$3.80) &
  54.55 ($\pm$4.36) \\
 \hline\hline
\textbf{CNN Posts} &
   2373 &
   1709 &
   13840 &
   633 &
  87.38 ($\pm$7.04) &
  58.13 ($\pm$12.02) &
  78.94 ($\pm$12.76) &
  66.96 ($\pm$12.37) \\
 \hline
\textbf{CNN Posts and Comments} &
   2304 &
   2068 &
   13481 &
   702 &
  85.07 ($\pm$1.25) &
  52.70 ($\pm$2.33) &
  76.65 ($\pm$3.45) &
  62.46 ($\pm$2.64) \\
 \hline\hline
\textbf{biLSTM Posts} &
   2297 &
   2495 &
   13054 &
   709 &
  82.73 ($\pm$8.11) &
  47.93 ($\pm$9.92) &
  76.41 ($\pm$10.94) &
  58.91 ($\pm$10.82) \\
 \hline
\textbf{biLSTM Posts and Comments}  &
   2288 &
   2370 &
   13179 &
   718 &
  83.36 ($\pm$2.27) &
  49.12 ($\pm$3.25) &
  76.11 ($\pm$3.86) &
  59.71 ($\pm$3.54) \\
 \hline
\end{tabular}
}
\caption{Summary of model performance}
\label{modelperformance}
\end{table}

The neural network models perform well as expected. 
Models such as XGBoost, Random Forests, and SVM had
disappointing performance, and a heterogeneous stacked classifier 
combining their predictions 
did not improve on the performance of the individual
predictors, suggesting that they make their errors on the same
records.

At first glance, the ANN models using posts perform better than those 
using posts and comments. However, the standard deviations of the performance
numbers show that the inclusion of comments provides stability for 
correctly identifying P\&D posts. 
The best performing model overall is CNN, especially with comments
included. Its precision is relatively low; 
of all the records that the model predicts to be P\&D, only 52.7\% are 
actually correct.  
If we look at the rate at which each class is predicted to be positive, 
a better outlook of the model is provided. 
Given a positive P\&D text, the model has a 76.65\% chance of classifying 
it correctly, whereas, if it is given a negative text, it has a 13.3\% 
chance of classifying it incorrectly as positive.
It is perhaps a little surprising
that biLSTM did not perform best since they are
typically strong predictors for natural language
problems.

\begin{figure}[p]
  \centering
    \includegraphics[width=0.8\textwidth]{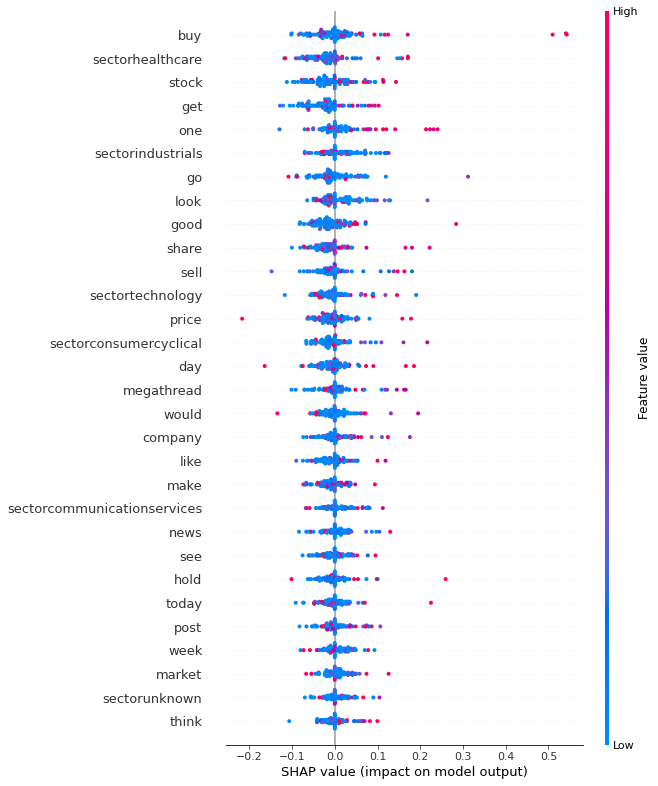}
    \caption{CNN SHAP Summary Plot for posts and comments}
    \label{cnn_pc}
\end{figure}

The SHAP Explainers produce diagrams that rank the attributes
by their impact on outcomes.
Figure~\ref{cnn_pc} shows the diagram for the CNN predictor for
posts and comments and the 30 most impactful words.
Although the influence of any single word is inevitably weak,
there are visible red dots to the right for many of these
words, indicating that higher frequencies of these words are
associated with P\&D events.
The names of the popular sectors
are indicator of P\&Ds, as are
words from the agreement model such as ``buy" and ``go".
Across the best performing models, the same set of words emerge as the 
most impactful features (not shown).

Misclassifications by the model have different impacts depending
on how and where it is used.
For an ordinary investor, a false positive (a post predicted to be
a P\&D when it isn't) means a missed opportunity for profit, but
a false negative means a financial loss.
For a regulatory body, a false positive is problematic, but a false
negative less so.
Table~\ref{misclassification} shows some of the examples
of misclassifications by the CNN model.

\begin{table}[t]
\centering
\resizebox{0.95\textwidth}{!}{
\begin{tabular}{|p{3cm}|p{3cm}|p{8cm}|}
\hline
\multicolumn{1}{|c|}{\textbf{Predicted Label}} & \multicolumn{1}{c|}{\textbf{Actual Label}} & \multicolumn{1}{c|}{\textbf{Misclassified Post}} \\ \hline\hline
       P\&D         &       Not P\&D       & sectorunknown about to soar                   \\ \hline
       P\&D          &       Not P\&D       & sectorunknown fitness equipment maker owner of bow flex completely sell out of most retail store how be this look just buy in share                   \\ \hline
       P\&D          &       Not P\&D       & quick all in sectorcommunicationservices pump my first time actually do something right the lambos go to be green for gain                   \\ \hline
       P\&D          &       Not P\&D       & blast off look like gold and oil will be big player this i also suggest look at sectortechnology                   \\ \hline
       P\&D          &       Not P\&D       & sectorenergy drop time to buy it be drop below which be its day low be it a good time to buy                   \\ \hline\hline
       Not P\&D         &       P\&D        & sectortechnology release patent news on thermal tech could be a mark sympathy play bust out over                   \\ \hline
       Not P\&D         &       P\&D        & sectorhealthcare do anyone understand why sectorhealthcare shoot up soo much i be not able to find any real catalyst                   \\ \hline
       Not P\&D         &       P\&D        & sectorhealthcare on the move this have potential reach today                   \\ \hline
       Not P\&D         &       P\&D        & sectorhealthcare to the moon                   \\ \hline
       Not P\&D         &       P\&D        & any thought on when to sell sectorenergy bought in late i be up after hour should i wait til tomorrow or sell as soon as possible in the am                   \\ \hline
\end{tabular}}
\caption{Examples of misclassified posts from CNN model}
\label{misclassification}
\end{table}

Some false positives, predicted to be P\&D from the text, but without
a corresponding market movement may be instances where the post
failed to attract enough attention to cause a measurable market
movement, or was so blatant that it was not credible to
typical investors.
Some false negatives may be because the posts were too short to
contain the required two words, because the pumping took
place on another platform or because a market
movement happened to match the timing of the post.

\section{Related work}

The application of data analytics for detecting market manipulation is 
a relatively new in the field of finance. 
Most research has focused on detecting trade-based manipulation
because it is most common \cite{wang_enhancing_2019}. 
Huang and Chang found that of the manipulation cases prosecuted in Taiwan 
from 1991 to 2010, 96.61\% were trade-based, 
and only 3.39\% were information-based \cite{Huang15}.
Some examples detecting trade-based manipulation are:
Ogut et al. \cite{ogut_detecting_2009} 
in the emerging Istanbul Stock Exchange,
Wang et al. \cite{wang_enhancing_2019}
for prosecuted manipulation cases reported 
by the China Securities Regulatory Commission, 
Cao et al. \cite{cao_detecting_2014}
using real trading data from four popular NASDAQ stocks with
synthetic cases of manipulation (spoofing and quote stuffing),
Cao et al. \cite{yi_cao_adaptive_2015} 
using seven popular NASDAQ and LSE stocks data 
injecting ten simulated stock price manipulations, 
Diaz et al. \cite{diaz_analysis_2011} 
using manipulation cases pursued 
by the U.S. Securities and Exchange Commission (SEC) in 2003,
and Golomohammadi et al. \cite{golmohammadi_detecting_2014} 
trying to detect three groups of 
manipulation schemes: marking the close, 
wash trades, and cornering the market. 

For information-based manipulation,
Victor and Hagemann \cite{victor_cryptocurrency_2019} 
looked at 149 confirmed P\&D schemes 
coordinated through Telegram chats and pumped via Twitter.
Using XGBoost, they built a model that achieved a sensitivity of 85\% 
and specificity of 99\%. 
They concluded that P\&Ds were frequent among cryptocurrencies that 
had a market capitalization of \$50 million or below 
and often involved trading volumes of several hundred thousand 
dollars within a short time-frame.

Mirtaheri et al. \cite{mirtaheri_identifying_2019} looked specifically at 
forecasting P\&Ds by combining the information from Twitter and Telegram.
They manually labelled known P\&D operation messages on Telegram, 
and then used SVMs with a stochastic gradient descent optimizer to 
label the remaining messages as P\&D or not. 
They used Random Forests to detect whether 
a manipulation event was going to take place within the market. 
Their results showed that they were able to detect, with reasonable accuracy,
whether there is an unfolding manipulation scheme occurring on Telegram. 
Their proposed model was able to achieve an accuracy of 87\% and an F1-Score of 90\%.

Some partially automated tools have also been developed.
These flag suspicious activities that can then by investigated
by regulators.
Delort et al. \cite{delort_2011} used Naive Bayes classifiers
to examine collected messages from HotCopper, 
an Australian stock message board.
They successfully identified messages of concern, but the number of
false positives was too high to use the model in an automated
way.
Owda et al. \cite{owda_financial_2017}
compared messages to lexicon templates of known illegal 
financial activities (e.g. Pump and Dump, Insider Information). 
They found that, of the 3000 comments that were collected on a daily basis, 
0.2\% were deemed suspicious.

\section{Conclusion}\label{ch:Conclusion}

The intersection of social media with low-cost trading platforms
and naive investors has made market manipulation an attractive
strategy. Pump\&dump is particularly simple to implement
since it requires only the dissemination of fictional information
about the future prospects for a stock.
This is particular easy for penny stocks where validating information
is difficult for ordinary investors, and where relatively small
purchase volumes can cause large price movements.

We investigate protecting investors, and assisting regulators,
by building predictive models that label social media posts (and
the responses they elicit) as potential drivers of P\&D events.
We do this by collecting posts and comments, developing a model
for a P\&D event based on patterns of price and volume
changes, using the match between posts and P\&D events
to label posts, and extending this labelling to comments using
an agreement model.
Natural language predictors then learn the language patterns
associated with P\&D manipulations, so that new manipulations
can be detected before they affect the market.

Data is imbalanced, since manipulations are rare, but
our best predictive model achieves an F1-score of 62\% and
an accuracy of 85\%.
Improvements in performance are limited by potential coincidences
between a post and a price and volume change that mimics
a P\&D, posts that fail to reach a sufficient audience to
cause the desired buying behaviour, and natural language issues
that arise from informal and short texts, and a specialised
vocabulary used in stock discussion forums.

\bibliography{thesis}

\begin{thebibliography}{10}

\bibitem{aggarwal_stock_2006}
R.K. Aggarwal and Guojun Wu.
\newblock Stock {Market} {Manipulations}.
\newblock {\em The Journal of Business}, 79(4):1915--1953, July 2006.

\bibitem{yfinance}
Ran Aroussi.
\newblock {yfinance}.
\newblock \url{https://aroussi.com/post/python-yahoo-finance}, Accessed:
  2021-09-02.

\bibitem{smote}
Kevin~W. Bowyer, Nitesh~V. Chawla, Lawrence~O. Hall, and W.~Philip Kegelmeyer.
\newblock {SMOTE:} {Synthetic Minority Over-sampling Technique}.
\newblock {\em CoRR}, abs/1106.1813, 2011.

\bibitem{brownlee_2020_lstm}
Jason Brownlee.
\newblock {A Gentle Introduction to Long Short-Term Memory Networks by the
  Experts}, Feb 2020.
\newblock
  \url{https://machinelearningmastery.com/gentle-introduction-long-short-term-memory-networks-experts/},
  Accessed: 2021-09-17.

\bibitem{brownlee_2020}
Jason Brownlee.
\newblock {A Gentle Introduction to XGBoost for Applied Machine Learning}, Apr
  2020.
\newblock
  \url{https://machinelearningmastery.com/gentle-introduction-xgboost-applied-machine-learning/},
  Accessed: 2021-09-15.

\bibitem{brownlee_2020_bilstm}
Jason Brownlee.
\newblock {How to Develop a Bidirectional LSTM For Sequence Classification in
  Python with Keras}, Aug 2020.
\newblock
  \url{https://machinelearningmastery.com/develop-bidirectional-lstm-sequence-classification-python-keras/},
  Accessed: 2021-09-18.

\bibitem{cao_detecting_2014}
Yi~Cao, Yuhua Li, Sonya Coleman, Ammar Belatreche, and T~M McGinnity.
\newblock Detecting {Price} {Manipulation} in the {Financial} {Market}.
\newblock In {\em Proceedings of the IEEE/IAFE Computational Intelligence for
  Financial Engineering (CIFEr)}, page~8, March 2014.

\bibitem{forde_2013}
Carole Comerton-Forde and Tālis~J. Putniņš.
\newblock {Stock Price Manipulation: Prevalence and Determinants}.
\newblock {\em Review of Finance}, 18(1):23--66, 03 2013.

\bibitem{csa_2020}
Canadian securities regulators warn public of coronavirus-related investment
  scams, Mar 2020.
\newblock \url{https://www.securities-administrators.ca/aboutcsa.aspx?id=1878},
  Accessed: 2021-09-06.

\bibitem{dacombe_2017}
James Dacombe.
\newblock {An introduction to Artificial Neural Networks (with example)}, Oct
  2017.
\newblock
  \url{https://medium.com/@jamesdacombe/an-introduction-to-artificial-neural-networks-with-example-ad459bb6941b},
  Accessed: 2021-09-16.

\bibitem{delort_2011}
Jean-Yves Delort, Bavani Arunasalam, and Cecile Paris.
\newblock Automatic moderation of online discussion sites.
\newblock {\em International Journal of Electronic Commerce}, 15(3):9--30,
  2011.

\bibitem{diaz_analysis_2011}
David Diaz, Babis Theodoulidis, and Pedro Sampaio.
\newblock Analysis of stock market manipulations using knowledge discovery
  techniques applied to intraday trade prices.
\newblock {\em Expert Systems with Applications}, 38(10):12757--12771,
  September 2011.

\bibitem{fast_empath_2016}
Ethan Fast, Binbin Chen, and Michael~S. Bernstein.
\newblock Empath: {Understanding} {Topic} {Signals} in {Large}-{Scale} {Text}.
\newblock In {\em Proceedings of the 2016 {CHI} {Conference} on {Human}
  {Factors} in {Computing} {Systems}}, pages 4647--4657, San Jose California
  USA, May 2016. ACM.

\bibitem{gandhi_2018}
Rohith Gandhi.
\newblock {Support Vector Machine - Introduction to Machine Learning
  Algorithms}, Jul 2018.
\newblock
  \url{https://towardsdatascience.com/support-vector-machine-introduction-to-machine-learning-algorithms-934a444fca47},
  Accessed: 2021-09-15.

\bibitem{ghanem-etal-2018-stance}
Bilal Ghanem, Paolo Rosso, and Francisco Rangel.
\newblock Stance detection in fake news a combined feature representation.
\newblock In {\em Proceedings of the First Workshop on Fact Extraction and
  {VER}ification ({FEVER})}, pages 66--71, Brussels, Belgium, November 2018.
  Association for Computational Linguistics.

\bibitem{golmohammadi_detecting_2014}
Koosha Golmohammadi, Osmar~R. Zaiane, and David Diaz.
\newblock {Detecting stock market manipulation using supervised learning
  algorithms}.
\newblock In {\em 2014 {International} {Conference} on {Data} {Science} and
  {Advanced} {Analytics} ({DSAA})}, pages 435--441, Shanghai, China, October
  2014. IEEE.

\bibitem{adasyn}
{Haibo He}, {Yang Bai}, E.~A. {Garcia}, and {Shutao Li}.
\newblock {ADASYN: Adaptive synthetic sampling approach for imbalanced
  learning}.
\newblock In {\em 2008 IEEE International Joint Conference on Neural Networks
  (IEEE World Congress on Computational Intelligence)}, pages 1322--1328, 2008.

\bibitem{Huang15}
Yu~Chuan Huang and Yao~Jen Cheng.
\newblock Stock manipulation and its effects: pump and dump versus
  stabilization.
\newblock {\em Review of Quantitative Finance and Accounting}, 44(4):791--815,
  May 2015.

\bibitem{kamps18}
Josh Kamps and Bennett Kleinberg.
\newblock To the moon: defining and detecting cryptocurrency pump and dumps.
\newblock {\em Crime Science}, 7(1):18, December 2018.

\bibitem{lecun_deep_2015}
Yann LeCun, Yoshua Bengio, and Geoffrey Hinton.
\newblock Deep learning.
\newblock {\em Nature}, 521(7553):436--444, May 2015.

\bibitem{li-caragea-2019-multi}
Yingjie Li and Cornelia Caragea.
\newblock Multi-task stance detection with sentiment and stance lexicons.
\newblock In {\em Proceedings of the 2019 Conference on Empirical Methods in
  Natural Language Processing and the 9th International Joint Conference on
  Natural Language Processing (EMNLP-IJCNLP)}, pages 6299--6305, Hong Kong,
  China, November 2019. Association for Computational Linguistics.

\bibitem{NIPS2017_7062}
Scott~M Lundberg and Su-In Lee.
\newblock {A Unified Approach to Interpreting Model Predictions}.
\newblock In I.~Guyon, U.~V. Luxburg, S.~Bengio, H.~Wallach, R.~Fergus,
  S.~Vishwanathan, and R.~Garnett, editors, {\em Advances in Neural Information
  Processing Systems 30}, pages 4765--4774. Curran Associates, Inc., 2017.

\bibitem{mirtaheri_identifying_2019}
Mehrnoosh Mirtaheri, Sami Abu-El-Haija, Fred Morstatter, Greg Ver~Steeg, and
  Aram Galstyan.
\newblock Identifying and {Analyzing} {Cryptocurrency} {Manipulations} in
  {Social} {Media}.
\newblock preprint, Open Science Framework, February 2019.

\bibitem{murphy_2020}
Chris~B. Murphy.
\newblock {Penny} {Stock}, Aug 2020.
\newblock \url{https://www.investopedia.com/terms/p/pennystock.asp}, Accessed:
  2021-09-02.

\bibitem{owda_financial_2017}
Majdi Owda, Pei~Shyuan Lee, and Keeley Crockett.
\newblock Financial {Discussion} {Boards} {Irregularities} {Detection} {System}
  ({FDBs}-{IDS}) using information extraction.
\newblock In {\em 2017 {Intelligent} {Systems} {Conference} ({IntelliSys})},
  pages 1078--1082, London, September 2017. IEEE.

\bibitem{rajaash}
Ashwin Rajadesingan and Huan Liu.
\newblock Identifying users with opposing opinions in {Twitter} debates.
\newblock In William~G. Kennedy, Nitin Agarwal, and Shanchieh~Jay Yang,
  editors, {\em Social Computing, Behavioral-Cultural Modeling and Prediction},
  pages 153--160, Cham, 2014. Springer International Publishing.

\bibitem{sabherwal_internet_2011}
Sanjiv Sabherwal, Salil~K. Sarkar, and Ying Zhang.
\newblock Do {Internet} {Stock} {Message} {Boards} {Influence} {Trading}?
  {Evidence} from {Heavily} {Discussed} {Stocks} with {No} {Fundamental}
  {News}: Do {Internet} {Stock} {Message} {Boards} {Influence} {Trading?}
\newblock {\em Journal of Business Finance \& Accounting}, 38(9-10):1209--1237,
  November 2011.

\bibitem{sec_2020}
{Investor Alerts and Bulletins: Frauds Targeting Main Street Investors --
  Investor Alert}, Apr 2020.
\newblock
  \url{https://www.sec.gov/oiea/investor-alerts-and-bulletins/ia_frauds},
  Accessed: 2020-09-06.

\bibitem{sec2_2020}
{Press Release: SEC Charges Microcap Fraud Scheme Participants Attempting to
  Capitalize on the COVID-19 Pandemic}, Jun 2020.
\newblock \url{https://www.sec.gov/news/press-release/2020-131}, Accessed:
  2020-09-06.

\bibitem{thomas-etal-2006-get}
Matt Thomas, Bo~Pang, and Lillian Lee.
\newblock Get out the vote: Determining support or opposition from
  congressional floor-debate transcripts.
\newblock In {\em Proceedings of the 2006 Conference on Empirical Methods in
  Natural Language Processing}, pages 327--335, Sydney, Australia, July 2006.
  Association for Computational Linguistics.

\bibitem{victor_cryptocurrency_2019}
Friedhelm Victor and Tanja Hagemann.
\newblock Cryptocurrency {Pump} and {Dump} {Schemes}: {Quantification} and
  {Detection}.
\newblock In {\em 2019 {International} {Conference} on {Data} {Mining}
  {Workshops} ({ICDMW})}, pages 244--251, Beijing, China, November 2019. IEEE.

\bibitem{wang_enhancing_2019}
Qili Wang, Wei Xu, Xinting Huang, and Kunlin Yang.
\newblock Enhancing intraday stock price manipulation detection by leveraging
  recurrent neural networks with ensemble learning.
\newblock {\em Neurocomputing}, 347:46--58, June 2019.

\bibitem{wild_2018}
Cody~Marie Wild.
\newblock {One} {Feature} {Attribution} {Method} to ({Supposedly}) {Rule}
  {Them} {All}: {Shapley} {Values}, Jan 2018.

\bibitem{wood_2020}
Thomas Wood.
\newblock {Random Forests}, Sep 2020.
\newblock
  \url{https://deepai.org/machine-learning-glossary-and-terms/random-forest},
  Accessed: 2021-09-16.

\bibitem{vidhya_2020}
{XGBoost Algorithm: XGBoost In Machine Learning}, May 2020.
\newblock
  \url{https://www.analyticsvidhya.com/blog/2018/09/an-end-to-end-guide-to-understand-the-math-behind-xgboost/},
  Accessed: 2021-09-15.

\bibitem{yi_cao_adaptive_2015}
{Yi Cao}, {Yuhua Li}, Sonya Coleman, Ammar Belatreche, and Thomas~Martin
  McGinnity.
\newblock Adaptive {Hidden} {Markov} {Model} {With} {Anomaly} {States} for
  {Price} {Manipulation} {Detection}.
\newblock {\em IEEE Transactions on Neural Networks and Learning Systems},
  26(2):318--330, February 2015.

\bibitem{yiu_2019}
Tony Yiu.
\newblock {Understanding Random Forest}, Aug 2019.
\newblock
  \url{https://towardsdatascience.com/understanding-random-forest-58381e0602d2},
  Accessed: 2021-09-16.

\bibitem{ogut_detecting_2009}
Hulisi Öğüt, M.~Mete~Doğanay, and Ramazan Aktaş.
\newblock Detecting stock-price manipulation in an emerging market: {The} case
  of {Turkey}.
\newblock {\em Expert Systems with Applications}, 36(9):11944--11949, November
  2009.

\end{thebibliography}
\end{document}